**Large Interferometer For Exoplanets (LIFE): VIII. Where is the phosphine? Observing exoplanetary $PH_3$ with a space based MIR nulling interferometer**

D. Angerhausen[1, 2, 3,*], M. Ottiger[1], F. Dannert[1], Y. Miguel[4, 5], C. Sousa-Silva[6], J. Kammerer[7], F. Menti[1], E. Alei[1, 2], B.S. Konrad[1, 2], H. S. Wang[1, 2], S.P. Quanz [1, 2], and the LIFE collaboration[8]

[1]ETH Zurich, Institute for Particle Physics & Astrophysics, Wolfgang-Pauli-Str. 27, 8093 Zurich, Switzerland
[2]National Center of Competence in Research PlanetS (www.nccr-planets.ch)
[3]Blue Marble Space Institute of Science, Seattle, United States
[4]SRON Netherlands Institute for Space Research , Sorbonnelaan 2, NL-3584 CA Utrecht, the Netherlands
[5]Leiden Observatory, University of Leiden, Niels Bohrweg 2, 2333CA Leiden, The Netherlands
[6]Center for Astrophysics | Harvard-Smithsonian, 60 Garden St, Cambridge, MA 02138, United States
[7]Space Telescope Science Institute, Baltimore, 3700 San Martin Dr, Baltimore, MD 21218, United States
[8]www.life-space-mission.com
*Correspondence: dangerhau@phys.ethz.ch

**Abstract**
Phosphine could be a key molecule in the understanding of exotic chemistry happening in (exo)planetary atmospheres. While it has been detected in the Solar System's giant planets, it has not been observed in exoplanets yet. In the exoplanetary context however it has been theorized as a potential biosignature molecule.  The goal of our study is to identify which illustrative science cases for $PH_3$ chemistry are observable with a space-based mid-infrared nulling interferometric observatory like the LIFE (Large Interferometer For Exoplanets) concept. We identified a representative set of scenarios for $PH_3$ detections in exoplanetary atmospheres varying over the whole dynamic range of the LIFE mission. We used chemical kinetics and radiative transfer calculations to produce forward models of these informative, prototypical observational cases for LIFEsim, our observation simulator software for LIFE. In a detailed, yet first order approximation it takes a mission like LIFE: (i) about 1h to find phosphine in a warm giant around a G star at 10 pc, (ii) about 10 h in $H_2$ or $CO_2$ dominated temperate super-Earths around M star hosts at 5 pc, (iii) and even in 100h it seems very unlikely that phosphine would be detectable in a Venus-Twin with extreme $PH_3$ concentrations at 5 pc. Phosphine in concentrations previously discussed in the literature is detectable in 2 out of the 3 cases and about an order of magnitude faster than comparable cases with JWST. We show that there is a significant number of objects accessible for these classes of observations. These results will be used to prioritize the parameter range for the next steps with more detailed retrieval simulations. They will also inform timely questions in the early design phase of a mission like LIFE and guide the community by providing easy-to-scale first estimates for a large part of detection space of such a mission.

# 1. Introduction

Phosphine is an essential molecule for our understanding of atmospheric chemistry in planets inside and outside our Solar System. It has been observed in the upper atmospheres of Jupiter and Saturn (e.g. Bregman et al. 1975; Larson et al. 1977; Weisstein & Serabyn 1996; Fletcher et al. 2009) and is also expected (not detected yet) to be similarly present in exoplanet atmospheres (e.g., Wang et al. 2017). If phosphine is indeed abundant in a wide range of (exo)planetary environments it is imperative to study its observability. Chemical kinetic models have demonstrated that phosphine can form in the deeper and hotter layers of giant planets, where the temperature and pressure are sufficiently high to make it the primary, thermochemically stable phosphorus-bearing species. Due to thermal disequilibrium, it is then dredged up to the upper layers of the atmosphere as a consequence of quenching and rapid vertical mixing and thus becomes observable (Visscher et al. 2006). This molecule is also expected to be the primary carrier of phosphorus in warm exoplanet atmospheres, where similar processes might happen. Even though the recent tentative detection of a trace of phosphine in the cloud decks of Venus has been challenged (Greaves et al. 2021, 2020a,b; Snellen et al. 2020; Encrenaz et al. 2020; Villanueva et al. 2020; Bains et al. 2020; Thompson 2021; Trompet et al. 2021; Lincowski et al. 2021; Akins et al. 2021), the interest of the community to this molecule, including its detectability in exoplanet atmospheres, has increased. First estimates in Sousa-Silva et al. (2020) show that $PH_3$ at ppm levels could be observed with the James Webb Space Telescope (JWST) using the technique of transit and eclipse spectrophotometry in $H_2$ and $CO_2$ dominated super-Earths orbiting even active late type stars at distances up to 5 pc. Wunderlich et al. (2020) discussed the detectability of $PH_3$ and other relevant potential biosignature gases for the case of the habitable-zone super-Earth LHS 1140 b with JWST and the Extremely Large Telescope (ELT). One of the major achievements of exoplanet research in the first half of this century will be the investigation of the atmospheric properties for a statistically significant number of temperate, terrestrial exoplanets (see e.g., Fujii et al. 2018). This is mostly driven by the search for habitable conditions and the opportunity to identify potential bio- or even technosignatures, based on a general understanding of the diversity of planetary bodies. Today, we are only at the beginning of understanding if there even are any "smoking gun" biosignatures nor are we certain about what numbers will make a statistically significant, representative data set of the planet atmosphere zoo. In the next two decades the first milestones on our roadmap will be taken with selected or proposed ground-based projects (see e.g., Snellen et al. 2013; Lovis et al. 2017; Ben-Ami et al. 2018) and space-based missions such as the Atmospheric Remote-sensing Infrared Exoplanet Large-survey (Ariel, Tinetti et al. 2018)[1], the Large UV/Optical/IR Surveyor (LUVOIR, The LUVOIR Team 2019)[2], the Habitable Exoplanet Observatory (HabEx, Gaudi et al. 2020)[3], and The Origins Space Telescope (OST, Meixner et al. 2019)[4]. These next generation observatories will likely deliver a first comprehensive and consistent, large observational set of exoplanetary atmospheres. A complementary approach to the most frequently discussed large space-based coronographic missions or the starshade concept is to separate the light emitted by the planet from that of its host star by means of an interferometer. LIFE is a project initiated in Europe with the goal to consolidate various efforts and define a roadmap that eventually leads to the launch of such a large, space-based mid-infrared (MIR) nulling interferometer (Quanz et al. 2018, 2019; LIFE collaboration et al. 2021). This mission should be able to investigate the atmospheric properties of a large sample of (primarily) temperate, terrestrial exoplanets. Centred around these clear and ambitious scientific objectives the project will define the relevant science and technical requirements. As a next step the status of key technologies will be re-assessed, and further technology development will be coordinated. For a detailed discussion of $PH_3$ in the context of more conventional biosignatures like the combination of oxygen and methane we refer to the standard literature (e.g., Des Marais et al. 2002; Seager et al. 2016; Schwieterman et al. 2018; Krissansen-Totton et al. 2018 and references therein.) as well as the LIFE papers (Alei et al. 2022, Konrad et al. 2022) addressing the detection of astrobiologically relevant species in Earth twins. A survey of the presence of phosphine in a large variety of exoplanetary contexts is needed to understand its potential as a biosignature molecule. If $PH_3$ is used as a biomarker in super-Earths planets with a $H_2$-dominated atmosphere we need to understand how the processes and chemical pathways work, i.e., the abiotic production of it in temperate (sub-)giants. As shown in figure 1, phosphine has a relatively unique signature in the MIR, that can be distinguished from other relevant species in this wavelength regime at the spectral resolutions provided by LIFE. The $O_3$, $NH_3$ and $PH_3$ bands cover a similar spectral range but feature significantly different shapes. However, it is still important to interpret these spectral signatures in the context of other system parameters and atmospheric constituents, when performing full spectral retrievals (see also section 4), that can for example derive if the atmosphere is oxidizing or not. Some combinations can only exist in different chemical regimes and would not interfere with one another. For example, PH3 is theorized to be present only in anaerobic environments (i.e., no $O_3$ present simultaneously). Here we present a first order study of the exploration space that is accessible to a mission like LIFE for an inventory study of phosphine in extrasolar planetary environments.

---

[1] https://arielmission.space/
[2] https://asd.gsfc.nasa.gov/luvoir/
[3] https://www.jpl.nasa.gov/habex/
[4] https://origins.ipac.caltech.edu/

## 2. Modelled Exoplanet Cases

*2.1. LIFEsim*

For the modelled observations presented here we use LIFEsim which simulates observations based on a detailed model of an interferometric exoplanet characterization observation, and accounts for noise from stellar leakage, local zodiacal, exo-zodiacal and dust emission, but no instrumental noise (see Dannert et al. 2022).

| *Parameter* | *Value* |
| --- | --- |
| Quantum efficiency | 0.7 |
| Throughput | 0.05 |
| Minimum Wavelength | 4 µm |
| Maximum Wavelength | 18.5 µm |
| Spectral Resolution | 50 |
| Imaging Baseline | 60-600 m |
| Nulling Baseline | 10-100 m |
| Apertures Diameter | 2m |
| Exozodi | 3x local zodi |

**Table 1.** *Overview of simulation parameters used in LIFEsim. These are the same standard values as e.g. used in Quanz et al. (2022)*

Following the same approach as Wang et al. (2017) and Sousa-Silva et al. (2020) in their analysis of the observability with JWST we used LIFEsim to produce synthetic observations of the outlined exoplanet cases with and without phosphine present in their atmospheres. For the presented output spectra, LIFEsim is configured with four apertures of 2 m diameter each, a broadband wavelength range of 4-18.5 µm and a spectral resolution of R = 50. We assume an exo-zodi level of 3 times the local zodi density (following the results from the HOSTS survey for the expected median level of emission in Ertel et al. 2020) and an interferometric baseline setup between 10 and 100 m (see Table 1).

## 2.2. Exoplanet Models

The main goal of this analysis was to identify the dynamic range of the LIFE observatory's exoplanet characterization potential by testing a few extreme, yet representative science cases. Fortunately for us the literature already contained studies of phosphine in hot to warm Jupiter sized planets (Wang et al. 2017) as well as various configurations of super-Earths (Sousa-Silva et al. 2020). In our scheme the Jupiter case represents one of the least challenging type of observations with LIFE, whereas the potentially habitable super-Earth case represents one of the "sweet spots" in terms of instrumental requirements as well as astrobiologically relevant observations. The additional scenario of the Venus twin was chosen to explore the extreme limits of an experiment like LIFE. The two cases from the literature we modelled with dedicated pipelines tailored to these setups, while the Venus case was modelled with a standard, open (online) tool. In the next subsections we describe in more detail these pipelines and refer to the original papers and the PSG documentation for all specifics. We decided for this approach since it would not have been trivial (or even impossible) to derive the all discussed cases from each from one or all of the used pipelines. E.g., the pipeline in Wang et al. (2017) was optimized for warm Jupiters while the methods in Sousa-Silva et al. (2020) focussed on Super-Earths. In Alei et al. 2022 we show that even in a full Bayesian retrieval using two different modelling pipelines has only minor impact on the retrieved abundances. Therefore, we assume that this will have little to no effect on the very general feasibility/detectability study we present here. For a more detailed study of the retrieval of chemical and physical properties of an Earth- twin planet that could be obtained with LIFE, we refer to Konrad et al. 2022. The characterization of the main features of an Earth-twin at various stages of its evolution (prebiotic, Great Oxygenation Event, modern atmosphere) is also discussed by Alei et al. 2022.

*2.2.1. Venus Twin*

Within the group of modelled science cases with the LIFE observatory concept, the detection and characterization of a Venus-like planet represents one of the most challenging observations. This case was chosen to explore the edge of the parameter space of possible detections with LIFE. Here we test if our local Solar System science case of potential phosphine in Venus would be detectable remotely. We use models from the Planetary Spectrum Generator (psg.gsfc.nasa.gov, Villanueva et al. (2018), PSG team priv. comms) with a *"standard Venus Template"* atmosphere and surface (see Table 2 for details) using the built-in *"volcanic cloud"* model.

We modelled various cases including 0, 5-20 ppb (the current estimate for Venus from Greaves et al. 2020c) and up to 310 ppm concentration (same as the super-Earth case in section 2.2.2 from Sousa-Silva et al. (2020), minimum concentration for remote JWST detection there). For the analysis here, we compare the detectability of the extreme 310 ppm scenario against one without any phosphine.

| Parameter | Value |
| --- | --- |
| Surface pressure | 92 bar |
| Molecular weight | 43.5 g/mol |
| Gases | $CO_2$, $N_2$, $CO$, $O_2$, $SO_2$, $H_2O$, $O_3$ |
| Phosphine | 0 vs 310ppm |
| Clouds | PSG "volcanic clouds" model |
| Surface temperature | 756 K |
| Surface gravity | 8.87 m/s$^2$ |
| Surface albedo | 0.22 |
| Diameter | 12104 km |
| Host star | Sun Twin at 10 pc |
| Distance to host | 0.72 AU |

**Table 2.** *Overview of simulation parameters used in the PSG model of a Venus twin.*

### 2.2.2. $H_2$- and CO2-dominated super-Earths around M star hosts

The models for the cases of $H_2$- and $CO_2$-dominated super-Earths around M star host are taken from Sousa-Silva et al. (2020, Table 3). The atmospheric composition for these scenarios is based on calculations with the photochemical model of Hu et al. (2012, for more details please see there and references therein). These model super-Earths with radii of 1.75 $R_\oplus$ and masses of 10 $M_\oplus$ are orbiting a 0.26 $R_S$ active 3000 K M-dwarf star at 5 pc distance. The assumed phosphine concentrations for the "with $PH_3$" cases are ($H_2$-dominated) 220 ppb and ($CO_2$-dominated) 310 ppm. To maintain equitable, temperate surface conditions (288 K), the H2- and CO2-dominated cases with no PH3 emissions are separated from their M dwarf host star by 0.042 and 0.034 AU, respectively. Sousa-Silva et al. (2020) used a photochemical model to explore how PH3 concentrations responded to two different scenarios: an 'active' M-dwarf UV flux model based on observations of GJ1214, and a 'quiet' M-dwarf for which the UV fluxes shortward of 300 nm are 0.1% of the 'active' scenario fluxes. The 'quiet' scenario resulted in PH3 accumulating to roughly 100 times higher concentrations for the same emission fluxes as compared to the 'active' M-dwarf scenario. The PH3 fluxes that would lead to the potential detection of PH3 are larger than the average terrestrial PH3 emission estimates but are comparable to the rates at which other biosignature gases are produced and well below observed site-specific PH3 fluxes on Earth (Sousa-Silva et al., 2020). The reader is directed to Sousa-Silva et al. (2020) for more details on the modeling and the interactions between phosphine and other atmospheric constituents.

| Parameters | $H_2$ dominated | $CO_2$ dominated |
| --- | --- | --- |
| Surface Temp. | 288 K | 288 K |
| Mass | 1.75 $R_\oplus$ | 1.75 $R_\oplus$ |
| Radius | 10 $M_\oplus$ | 10 $M_\oplus$ |
| Phosphine | 0 - 220 ppb | 0 - 310 ppm |
| Host star | M star at 5 pc | M star at 5 pc |
| Host star Temp. | 3000 K | 3000 K |
| Distance to star | 0.042 AU | 0.034 AU |

**Table 3.** *Overview of simulation parameters used in the Super Earth models. For more details, see (Sousa-Silva et al. 2020).*

### 2.2.3. Warm Giant Planets around G Type Hosts

We use models from Wang et al. (2017, Table 4) for a Jupiter-like planet around a G-type star using chemical kinetics to calculate the abundances of different species, which were then input to the Smithsonian Astrophysical Observatory 1998 (SAO98) radiative transfer code (Traub & Stier 1976; Kaltenegger & Traub 2009) to calculate the transmission and emission spectra. Their chemical calculations show that $PH_3$ is the main phosphorous carrier for exoplanets between 400 and 2000K.

In these models, the effects of vertical mixing are considered but those of photochemistry in the upper atmosphere of these planets are neglected. The rationale for this is that while photochemistry is relevant in very hot planets with a hydrogen dominated atmosphere of solar composition close to their stars, their effect on warm planets is limited (e.g., Miguel & Kaltenegger 2014); in addition, vertical mixing effects are the most relevant ones in the pressures probed by transmission spectroscopy (at the millibar level). Wang et al. (2017) used the solar elemental (atomic) abundances of C, H, O, N, S and P (Asplund et al. 2009) and let the code calculate the molecules that will be formed given the conditions in the planetary atmosphere calculation. We note that the Wang et al. (2017) calculation was done without considering the effect of photochemistry which might affect abundances in the upper parts of the atmosphere (pressures< $1e^{-4}$ bar) but will most likely have a small effect on the detectability of different species, in particular in MIR emission (vs e.g., transit transmission in the optical, NIR).

| *Parameter* | *Value* |
| --- | --- |
| Temperature | 500 K |
| Radius | 1 $M_J$ |
| Mass | 1 $R_J$ |
| Phosphine | 0 vs solar |
| Host star | G-star at 10 pc |
| Distance to host star | 0.31 AU |

**Table 4.** *Overview of model parameters for case 3. "Solar" abundances refer to Asplund et al. (2009).*

## 3. Results

*3.1. Estimated yields*

In a first step of our study, we checked the availability of potential targets by analysing the occurrence of the modelled cases in the Solar neighbourhood accessible for the theorised observations. Therefore, we checked how many of these potential targets are already known and calculated how many of these LIFE would be able to detect in its detection phase alone even if no other methods would have found those by the time of LIFE's launch.

For the yield simulations, sections of the parameter space corresponding to the described cases are selected. These restrictions to the parameter space are shown in Table 5. We used planet occurrence rates inferred from NASA's Kepler mission to estimate the total expected number of exoplanets in the solar neighbourhood and computed the fraction of these, which is detectable with LIFE within 10 h of integration time. For this task, we create synthetic exoplanet populations using the NASA ExoPAG SAG13 occurrence rates (Kopparapu et al. 2018) for all spectral types with P-pop[5] (Kammerer & Quanz 2018) and select different sub-samples of planets (Venus twins, super-Earths, and warm Jovians) according to Table 5. For more details on the yield calculations, see Quanz et al. (2022).

| Parameter | Venus twin | super-Earth | Warm Jovian |
|---|---|---|---|
| Surface Temp. | $0.5\,R_\oplus < R < 1\,R_\oplus$ | $1\,R_\oplus < R < 1.75\,R_\oplus$ | $6\,R_\oplus < R < 14.3\,R_\oplus$ |
| Eq. Temperature | 200–800 K | 200–800 K (a) | 200–800 K |

**Table 5.** *Restrictions in parameter space for exoplanet categories (a) For the eHZ cases we defined it by the "Recent Venus" and "Early Mars" scenarios for $1\,M_\oplus$ planets from Kopparapu et al. (2014).*

For each of these sub-samples, we then compute the number of planets which could be detected with an SNR > 7 in 10 h of integration time using LIFEsim (Dannert et al. 2022, we require an SNR > 7 to leave a margin for additional instrumental noise). Table 6 shows planets within 20 pc around FGK stars and Table 7 shows planets within 10 pc around M stars. We limit the M star case to within 10 pc because the survey efficiency (number of detections per observed target) becomes low further out (see Quanz et al. 2022). For the case of the super Earth around an M star at 5 pc, the on-sky separation between the planet and its host star lies at 6.8 mas. The detection and characterization of the planet at this close proximity of about 0.3 lambda/$B_{nulling}$ in terms of the interferometric array properties illustrates the capability of LIFE to directly image close-in targets. This is made possible by the suppression of the central null being a continuous function down to the line-of-sight, which results in the inner working angle of the observatory being a strong function of the target planet luminosity at a fraction of lambda/$B_{nulling}$. It should be noted that an absolute lower bound of the inner working angle is given by the fringe spacing due to the imaging baseline, i.e., 0.5 lambda/$B_{imaging}$.

---

[5] https://github.com/kammerje/P-pop

| Case | Known | Total expected | LIFE detectable |
|---|---|---|---|
| Venus Twin | 1 | 247.6 ± 15.7 | 8.1 ± 3.0 |
| Super-Earth | 2 | 166.8 ± 12.9 | 44.9 ± 6.9 |
| eHZ Super-Earth | 0 | 31.7 ± 5.6 | $0.5^{+0.7}_{-0.5}$ |
| Warm Jovian | 13 | 12.2 ± 3.5 | 11.9 ± 3.4 |

*Table 6.* *Number of potential targets in the respective cases within 20 pc for FGK host stars. Column one shows the planets already known from the NASA Exoplanet Archive, column two the number of planets expected from Kepler statistics (using P-pop with the SAG13 occurrence rates), and column three the number of planets detectable with LIFE within 10 h of integration time.*

The limits of the "boxes" shown here are taken to be wider than the cases described in our examples and are chosen to be aligned with Kopporappu et al. 2016. The point we are making is that there will be dozens of planets in this parameter space to explore that will allow us to derive the presence or non-presence of phosphine within the order of integrations times used in our examples.

| Case | Known | Total expected | LIFE detectable |
|---|---|---|---|
| Venus Twin | 0 | 78.7 ± 8.9 | 18.5 ± 4.4 |
| Super-Earth | 13 | 53.9 ± 7.3 | 31.9 ± 6.0 |
| eHZ Super-Earth | 4 | 14.0 ± 3.7 | 4.5 ± 2.0 |
| Warm Jovian | 1 | 1.7 ± 1.3 | 1.6 ± 1.3 |

**Table 7.** *Same as Table 6 but for M star hosts within 10 pc only.*

### 3.2. Venus Twin

Figure 2 shows the results of a simulated observation of a Venus twin around a Sun twin at 10 pc. We compare a simulated 100h observation of an atmosphere with 310 ppm of phosphine to a phosphine free model. The significance is only ~ 0.15 sigma and the phosphine is thus not detectable.

### 3.3. $H_2$- and $CO_2$-dominated super-Earth around M star host

Figures 3 and 4 show the results of simulated observations of $H_2$- and $CO_2$-dominated super-Earths orbiting an M star host at 5 pc (from Sousa-Silva et al. (2020), see section 2.2.2). We modelled a 10h observation of these atmospheres with 220 ppb and 310 ppm of phosphine respectively and compared them to phosphine free models for the same cases. We get a solid detection of up to 5 σ per spectral channel in the bands between 8 and 12 µm that show the biggest difference between a $PH_3$ bearing atmosphere and one without.

### 3.4. 500 K Giant around G star host

Figure 5 shows the results of a simulated observation of a 500 K giant planet orbiting at G star at 10 pc. We get a very solid detection of up to 50 σ per spectral channel in the band around 4.5 µm and up to 20 σ per spectral channel in the bands between 8 and 12 µm that show the biggest difference between a $PH_3$ bearing atmosphere and one without.

## 4. Discussion and conclusions

By applying a detailed mission and instrument simulator for a future mid-infrared space interferometer to map the detectability space of phosphine in three different, extreme planet types and scenarios we find that:

1. Observations of extrasolar $PH_3$ in a Venus-like planet around a G-type stars is very challenging. Even after 100 hours of simulated observation the difference in two extreme forward models (no vs 310 ppm of $PH_3$) phosphine is not detectable.

2. For the cases of $H_2$ and $CO_2$ dominated super-Earths around M star hosts a mission like LIFE can produce highly informative observations in a relatively small amount of time. Applied to a suite of models for these planets taken from the literature, LIFEsim confirms that phosphine concentrations of 220 ppb ($H_2$ case) and 310 ppm ($CO_2$ case) can clearly be distinguished from $PH_3$ free atmospheres with integration times in the order of 10 hours.

3. Giant planets, such as the example of the 500 K Jupiter can quickly and easily be characterized in the order of one hour. LIFE will allow us to make - if present - detections of phosphine in many giant exoplanets and help understand its relevance and occurrence.

In Table 8 we show a comparison of the observation times needed to get a significant detection of phosphine in the described cases. In comparison to the studies of JWST observability of the same scenarios at comparable signal to noise ratios in Sousa-Silva et al. (2020) and Wang et al. (2017), we can report that a concept like LIFE will perform more than an order of magnitude faster: the super-Earths scenarios take 131 hours (for a $H_2$-dominated atmosphere) and 48 hours (for a $CO_2$- dominated atmosphere) of JWST secondary eclipse emission spectrophotometry Sousa-Silva et al. (2020) in comparison to about 10 hours of observation time needed with LIFE. While the warm Jupiter scenario in Wang et al. (2017) is observable only in transit transmission spectrophotometry within about 30 hours of time with JWST, it takes only 1h with LIFE for a solid detection of phosphine. It is also important to note that the JWST observations require multiple visits to add up the times from several eclipse/transit events while the LIFE observation can be done in a single visit and on non-transiting systems as well.

| Observation type | 500K Giant | $H_2$ Super Earth | $CO_2$ Super Earth |
|---|---|---|---|
| JWST transit/transmission | ~29h | ~91h | >200h |
| JWST eclipse/emission | >30h | ~131h | ~48h |
| LIFE nulling interferometry/emission | <1h | <10h | <10h |

**Table 8.** *Overview of observation times needed to detect the modelled levels of $PH_3$. JWST numbers from Wang et al. (2017) and Sousa-Silva et al. (2020).*

A caveat of the presented analysis is that it was done for presumably extreme cases and for the comparison of forward models with and without phosphine in the respective atmospheres. Therefore, the next steps for us will be to simulate more examples and run full Bayesian retrievals to get to more quantitative assessments also on the limits of phosphine detectability. The analysis presented here will help to prioritize the parameter range to go for in the next deeper dives into these atmospheres with more detailed retrieval simulations. These results will inform crucial questions and decisions in the early design phase of a mission like LIFE. In combination with other early results of the LIFE initiative these results will help to get a better idea on how to handle the overall planet statistics in the planning phase of a large future missions for terrestrial exoplanet characterization by giving a first order estimate of the realistically explorable parameter space. To build upon these estimates the LIFE team is currently working on a comprehensive simulation of the consequences on scheduling for the survey and characterisation phases of the LIFE mission. At this end we are also investigating novel machine learning methods that are necessary to expand and augment our front-to-end simulations of the full LIFE survey, including full retrievals (see e.g., Gebhard et al. 2022). These results will deliver timely information needed for the early definition of sensitivity, wavelength coverage and spectral resolution requirements on the technology side and set important limits for the modellers to reduce the computing load for upcoming future work such as calculations of spectral/photochemical grids.

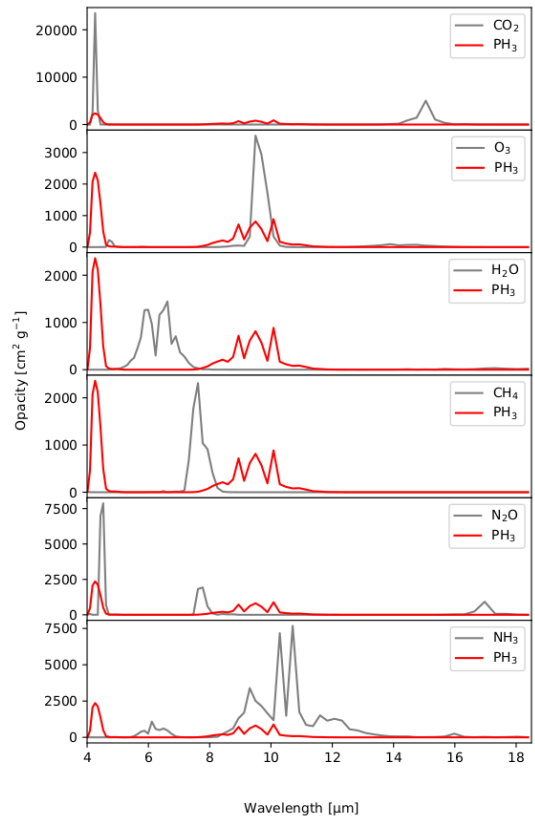

*Figure 1.* Comparison of the opacities of phosphine with other relevant species in the mid infrared. While some other species are optically active at similar wavelengths, it has a very distinct shape that help to distinguish it.

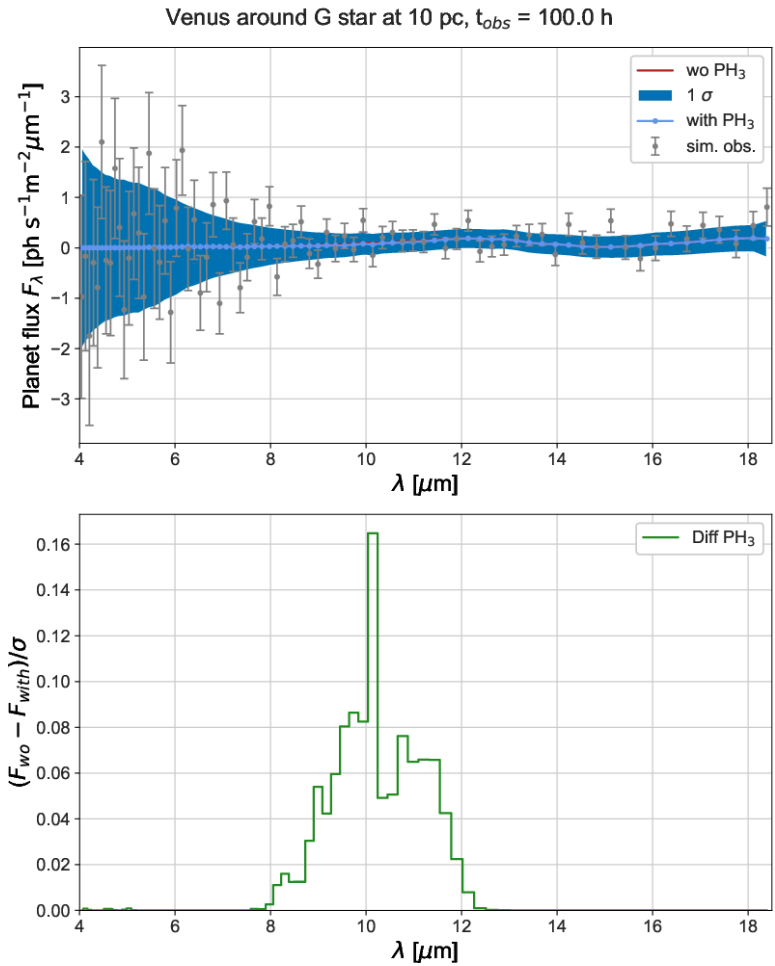

*Figure 2.* Detectability of phosphine in the emission spectrum of a Venus Twin planet, after 100 hours of observation with LIFE. Top: planet flux for atmospheres with and without $PH_3$. The blue area represents the 1-σ sensitivity; the grey error bars show an individual simulated observation. Bottom: Statistical significance of the detected difference between an atmospheric model with and without phosphine.

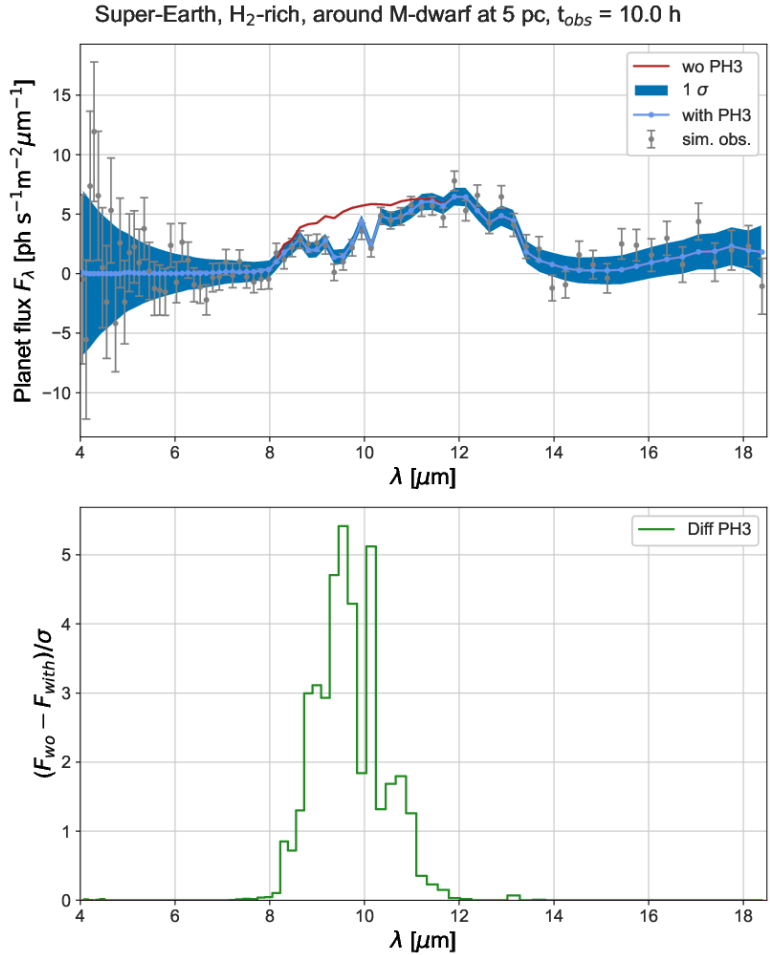

*Figure 3.* Detectability of phosphine in the emission spectrum of a super Earth exoplanet (10M$_\oplus$; 1:75R$_\oplus$) with a hydrogen-rich atmosphere orbiting an active M-dwarf, after 10 hours of observation with LIFE. Top: planet flux for atmospheres with and without PH$_3$, respectively; blue and red curves represent a modelled atmosphere with a PH$_3$ mixing ratio of 310 ppm and an atmosphere without PH$_3$. The blue area represents the 1-$\sigma$ sensitivity; the grey error bars show an individual simulated observation. Bottom: Statistical significance of the detected difference between an atmospheric model with and without phosphine.

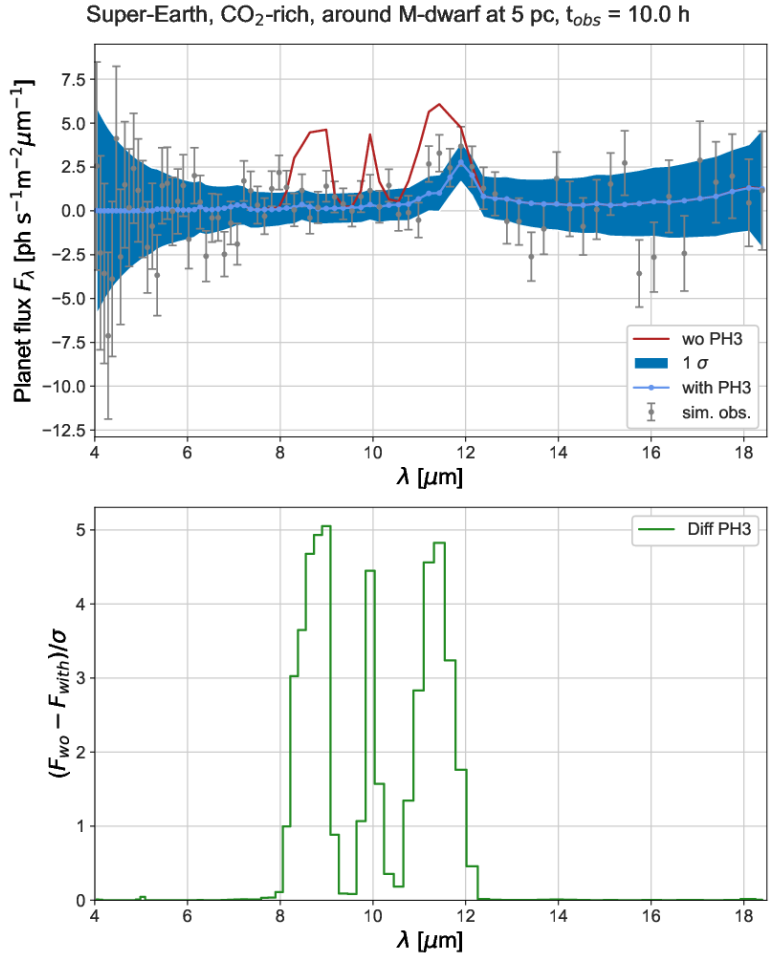

*Figure 4.* Detectability of phosphine in the emission spectrum of a super Earth exoplanet (10$M_⊕$; 1:75$R_⊕$) with a carbon dioxide-rich atmosphere orbiting an active M-dwarf, after 10 hours of observation with LIFE. Top: planet flux for atmospheres with and without $PH_3$, respectively; blue and red curves represent a modelled atmosphere with a $PH_3$ mixing ratio of 310 ppm and an atmosphere without $PH_3$. The blue area represents the 1-$σ$ sensitivity; the grey error bars show an individual simulated observation. Bottom: Statistical significance of the detected difference between an atmospheric model with and without phosphine.

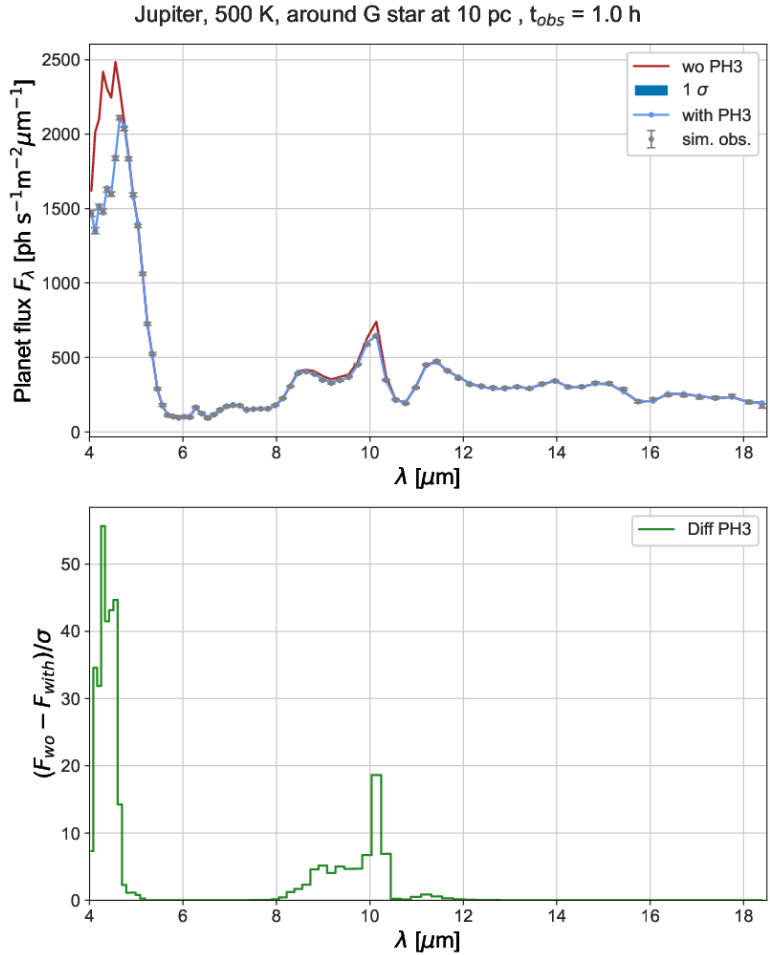

*Figure 5.* Detectability of phosphine in the emission spectrum of 500 K warm Jupiter like planet, after 1 hour of observation with LIFE. Top: planet flux for atmospheres with and without $PH_3$. The blue area represents the 1-$\sigma$ sensitivity; the grey error bars show an individual simulated observation. Bottom: Statistical significance of the detected difference between an atmospheric model with and without phosphine.

**Acknowledgements**
Part of this work was supported by SNF, NCCR PlanetS and the Heising-Simons Foundation. The authors thank the referees as well as the participants of AbSciCon 2022 for very useful feedback on this work.

**CRediT author statement**

**DA**: Conceptualization**,** Methodology, Data Curation, Writing - Original Draft, Writing – Review & Editing, Visualization, Supervision, Project administration; **MO:** Software, Methodology, Visualization; **FD:** Software, Methodology, Formal analysis, Data Curation; **YM:** Methodology, Writing - Review & Editing; **CSS:** Methodology , Writing - Review & Editing; **JK:** Methodology, Software, Formal analysis, Writing - Review & Editing; **FM:** Writing - Review & Editing; **EA:** Writing - Review & Editing, Visualization , **BK:** Writing - Review & Editing; **HW:** Writing - Review & Editing; **SQ:** Writing - Review & Editing, Supervision